\documentclass[journal,article,submit,moreauthors,pdftex,10pt,a4paper]{Definitions/mdpi}

\newcommand{\be}{\begin{equation}}
\newcommand{\ee}{\end{equation}}
\newcommand{\bea}{\begin{eqnarray}}  
\newcommand{\eea}{\end{eqnarray}}

\usepackage{graphicx}
\usepackage{dcolumn}
\usepackage{bm}

\usepackage{verbatim}
\usepackage{amsmath}
\usepackage{amssymb} 
\usepackage{color}
\usepackage{multirow}  

\usepackage{soul}

\firstpage{1} 
\pubvolume{xx}
\issuenum{1}
\articlenumber{1}
\pubyear{2024}
\copyrightyear{2024}
\externaleditor{Academic Editors: Stephen Xsu, Xavier Calmet, Roberto Casadio} 
\history{Received: date; Accepted: date; Published: date}

\Title{Look beyond additivity and extensivity of entropy for black hole and cosmological horizons. }

\Author{Mariusz P. D\c{a}browski $^{1,2,3}$}

\address{%
$^{1}$ \quad Institute of Physics, University of Szczecin, Wielkopolska 15, 70-451 Szczecin, Poland; Mariusz.Dabrowski@usz.edu.pl\\
$^{2}$ \quad National Centre for Nuclear Research, Andrzeja So{\l}tana 7, 05-400 Otwock, Poland; \\
$^{3}$ \quad  Copernicus Center for Interdisciplinary Studies, Szczepa\'nska 1/5, 31-011 Krak\'ow, Poland}  

\abstract{We present a comparative analysis of the plethora of nonextensive and/or nonadditive entropies which go beyond the standard Boltzmann-Gibbs formulation. After defining the basic notions of additivity, extensivity, and composability, we discuss the properties of these entropies and their mutual relations, if they exist. The results are presented in two informative tables supposedly of strong interest to gravity and cosmology community in the context of intensively explored recent days the horizon entropies for black hole and cosmological models. This is since gravitational systems admit long-range interactions which usually lead to a break of the standard additivity rule for thermodynamical systems composed of subsystems in Boltzmann-Gibbs thermodynamics. The features of additivity, extensivity, and the composability are listed systematically. Some brief discussion on the validity of the notion of equilibrium temperature for nonextensive systems is also presented. }

\keyword{thermodynamics, entropy, additivity, extensivity, long-range interactions, horizons.}

\begin{document}


\section{Introduction}

It is widely known that the Boltzmann-Gibbs thermodynamics (from now on BG) and statistical mechanics are additive and extensive \cite{Swendsen}. The core physical quantity which refers to these theories is the entropy, which is assumed to be extensive, since it relates to the negligence of the long-range forces between thermodynamic subsystems. This assumption is justified only, when the size of the system exceeds the range of the interaction between its components. As a result, the total entropy of a composite system is equal to the sum of the entropies of the individual subsystems (additivity) and the entropy grows with the size of the system or its configuration space (extensivity). 


However, contemporary physics exhibits a number of systems for which the long-range forces are important. The very examples of such systems are gravitational systems since gravity is long-range interactive, and besides, it is strongly non-linear when its extreme regimes are taken into account. Strong gravity characterises all the compact astrophysical objects in the Universe like white dwarfs, neutron and boson stars, quark stars etc., with the most extreme and most intriguing - the black holes. The latter are surrounded by the horizons which areas, according to Bekenstein and Hawking \cite{Bekenstein:1973ur,Hawking:1975vcx}, can be interpreted as the entropy, and so it is possible to formulate the appropriate laws of thermodynamics. Since for the black holes the Bekenstein entropy {\it scales with the area} and not with the volume (size), it is consequently a nonextensive quantity \cite{Tsallis:2012js,Biro:2013cra,Czinner:2017bwc,Czinner:2015eyk,Czinner:2017tjq,Tsallis:2019giw}. In addition to that,  because of a long-range interaction nature of gravity, the Bekenstein entropy is also nonadditive. 
 
In fact, a number of nonadditive and/or nonextensive entropies have been proposed in the literature \cite{Tsallis:1987eu,Renyi1,SM,sharma1975new,Kaniadakis:2002zz,Kaniadakis:2005zk,Barrow:2020tzx}. Most of them have been applied to gravitational systems both in astrophysics and in cosmology  and there is a debate, if they can serve dark energy \cite{Saridakis:2020zol,MPD_Salz_2020,Saridakis:2020lrg,AlMamon:2020usb,Asghari:2021lzu,Adhikary:2021xym,Nojiri:2021czz,Drepanou:2021jiv,DiGennaro:2022ykp,DiGennaro:2022grw,Nojiri:2021jxf,Anagnostopoulos:2020ctz,Sheykhi:2018dpn,FeiziMangoudehi:2022rwj,Ghoshal:2021ief,Luciano2022a,Leon:2021wyx,Almada2022_Kaniad_test}, which is  
specifically called the {\it holographic dark energy} \cite{Wang:2016och}. Amazingly, a number of these explorations do not acknowledge the fact of nonextensivity of the systems, which makes some issues related to their firm thermodynamical background \cite{PhysRevE.67.036114}. All this motivates us to investigate the problem of nonextensive entropies applications in some gravitational framework. 

In this paper, we explore the topic of additivity and extensivity of entropies which goes beyond the standard BG thermodynamics, being strongly motivated by gravitational interaction. Our focus is put on non-standard, but better fitting to gravitational systems entropies, such as: Bekenstein entropy  \cite{Bekenstein:1973ur,Hawking:1975vcx}, Tsallis $q$-entropy \cite{Tsallis:1987eu,TSALLIS1998534}, Tsallis-Cirto $\delta$-entropy \cite{Tsallis:2012js}, Barrow $\Delta$-entropy \cite{Barrow:2020tzx}, Tsallis $q, \delta$ entropy, Tsallis-Jensen $q, \gamma$-entropy \cite{Tsallis2024}, R\'enyi entropy \cite{Renyi1}, Landsberg $U$-entropy \cite{soton29487}, 
Sharma-Mittal entropy  \cite{SM,sharma1975new} , and Kaniadakis entropy \cite{Kaniadakis:2002zz,Kaniadakis:2005zk} . 

The following is the outline of the paper. In Sec. \ref{BGT}, we define additivity and extensivity in thermodynamical systems and try to establish some generalities about possible composition rules for the entropy. In Sec. \ref{BeyondBGT} we go beyond the definitions of additivity and extensivity. In Sec. \ref{zoo} we constructively review and compare the plethora of nonadditive and/or nonextensive entropies together with accompanying nonadditive and/or nonextensive thermodynamical quantities. Then, we make the classification of the entropies under study with respect to additivity and extensivity properties as well as with the application of the appropriate composition rules. Finally, in Sec. \ref{summary}, we summarize the paper.

\section{Boltzmann-Gibbs thermodynamics and statistical mechanics}
\label{BGT}
Boltzmann-Gibbs thermodynamics and statistical mechanics are based on two key hypotheses which are that the entropy is extensive and that the internal energy and entropy follow the additive composition rule for the system made of some subsystems. All physical relations in BG statistical mechanics are defined in light of these conditions which, in fact, rely on ignoring long-range forces between thermodynamical subsystems.

The Boltzmann-Gibbs (BG) entropy is defined as \cite{Swendsen}
\be
S_{BG}=-k_B\sum_{i=1}^n p_i \ln p_i = k_B \sum_{i=1}^n p_i \ln {\frac{1}{p_i}} ,
 \label{BG}
\ee
where $p_i$ is the probability distribution defined on a configuration space $\Omega$ with the number of degrees of freedom (states) $n$, $k_B$ is the Boltzmann constant, and the condition that the total probability must be equal to one $\sum p_i = 1$ is fulfilled. For the case of all probabilities equal, i.e. for $p_i =$ const. $= p$, we get 
\be
\sum_{i=1}^{n} p_i = 1 = np \hspace{0.2cm} \Rightarrow p = 1/n .
\label{p_i_equal}
\ee
After applying (\ref{p_i_equal}) to (\ref{BG}), one obtains that 
\be
S_{BG} = k_B \ln{n} ,
\label{BGequalp}
\ee
which means that the entropy is proportional to the number of states $n$ in the configuration space $\Omega$. 

In view of the key properties of BG thermodynamics, and in the context of our investigations beyond these properties, we can define additivity and extensivity in quite a general way following some literature \cite{soton29487,Swendsen2011,Mannaerts_2014} as below. 

\subsection{Additivity} 

Additivity means that for a given physical or thermodynamical quantity $f$, the following composition rule is fulfilled: 
\be
f(A+B) = f(p_{A \cup B}) = f(p_A p_B) = f(p_A) + f(p_B) =  f(A) + f(B) ,
\label{addrule}
\ee
where A, B are independent subsystems, equipped with the sets of configuration space degrees of freedom $\Omega_A$ and $\Omega_B$, and corresponding probabilities $p_A$ and $p_B$. The composite system $A \cup B$ allows the probability $p_{A \cup B}$ and it is equipped with the set of configuration space degrees of freedom $\Omega_{A \cup B}$. If the subsystems $A$ and $B$ are assumed to be independent, then it happens that the probabilities are related by $p_{A \cup B}=p_A p_B$, which allows the transition leading to the additivity rule (\ref{addrule}) \cite{Cimdiker:2022ics}.

If a particular case of the entropy $S$ is taken into account, then (\ref{addrule}) reads 
\be
S(A + B) = S(A) + S(B)  .
\label{addSrule}
\ee

\subsection{Extensivity} 

Let us assume that there is a set of physical quantities $(X_0, X_1, X_2, \ldots, X_k)$ such that $X_0 = f(X_1, X_2, \ldots, X_k)$. Extensivity of a selected physical quantity means that the function $f$ which describes this quantity is {\it homogeneous degree one}  \cite{soton29487,Swendsen,Swendsen2011} i.e. that 
\be
f(aX_1,aX_2,...,aX_k)=af(X_1,X_2,...,X_k)
\label{homog}
\ee
for every positive real number $a>0$, for all $X_1, X_2, ...X_k$. Taking $k=3$, so that we have only four quantities $X_0, X_1, X_2, X_3$, and assuming that they are the entropy $S$, the energy $E$, the volume $V$, and the mole number $N$ accordingly, we can obtain the standard Boltzmann-Gibbs thermodynamical extensivity relation for the entropy \cite{soton29487}
\be
S(aE,aV,aN) = aS(E,V,N) .
\label{homogS}
\ee
 In fact, the property (\ref{homog}) is called 'homogeneity', and is considered the most general definition of extensivity (cf. \cite{soton29487}). 

In standard textbooks of thermodynamics, one commonly uses less general definition of an extensive quantity, which says that if a system's total number of states in the configuration space $\Omega$ is proportional to its number of degrees of freedom, then this quantity (such as the entropy, for example) is extensive. For BG entropy, as we have shown in (\ref{BGequalp}), one has that $S_{BG}(n)=k_B \ln{ (n)} \propto n$, where $n$ is the total number of states in the system. 

 The advantage of definition (\ref{homog}) is that one does not refer to any kind of geometrical or bulk properties of a system such as the 'size', though the geometrical size of a system seems intuitively to be related to the number of states or degrees of freedom. 

 \subsection{Concavity}
 
 Concavity is the feature of the functions which read as \cite{soton29487,Tsallis:2012js}
 \be
 f(ax + (1-a)y) \geq af(x) + (1-a)f(y) \hspace{0.3cm} (a>0).
 \label{concavity}
 \ee
 In the context of thermodynamics, concavity of entropy guarantees that the system in thermodynamic equilibrium is {\it stable}. 

\section{Beyond Boltzmann-Gibbs thermodynamics}
\label{BeyondBGT}

\subsection{Composability} 

Let us consider two independent systems $A$ and $B$, combined as a single Cartesian product $A \times B$ of the states of $A$ and $B$ with the requirement that \cite{Tsallis2024}: 
\be
S(A \times B, \Upsilon) = k_B g \left(\frac{S(A)}{k_B}, \frac{S(B)}{k_B} \right) ,
\label{Composable}
\ee
where $g$ is a smooth function of $S(A)$ and $S(B)$ and $\Upsilon$  is a parameter, which in the limit $\Upsilon \to 0$ give an additive composition rule (\ref{addSrule}). If the systems $A$ and $B$ fulfil the condition (\ref{Composable}), then their combined system $A + B$ is called {\it composable}. Of course, the BG system is composable. 

\subsection{Beyond additivity} 


Additivity is violated, if the rule (\ref{addSrule}) does not hold. In such a case, one can have two options \cite{soton29487}. The first one is when 
\be
S(A+B) \geq  S(A) + S(B) ,
\label{superrule}
\ee
which is called {\it superadditivity}, and it leads to the tendency of the system to clump its pieces/subsystems. The second one is when
\be
S(A+B) <  S(A) + S(B) ,
\label{subrule}
\ee  
which is {\it subadditivity}, and it tends to fragment the system into pieces, rather than clump. A cosmological similarity of such a system is phantom \cite{Caldwell2002,Caldwell2003}, since it splits spontaneously into pieces under (anti)gravity beginning with the largest size pieces and terminating at the smallest \cite{ Dabrowski_EPJ_2015}. 

In the literature, there are a number composition rules for nonadditive systems, which we introduce in Section \ref{zoo}. One of them, which generalizes the additive composition rule (\ref{addSrule}) into a nonadditive case, is the Ab\'e rule \cite{Abe_2001_PLA,Abe_2001a,Abe_2001c}. It fulfils the composability requirement given by (\ref{Composable}). If applied to entropy, it reads as follows
\be
S(A+B) =  S(A) + S(B) + \frac{\Upsilon}{k_B} S(A) S(B),
\label{Aberule}
\ee
where $\Upsilon$ takes numerical values according to the statistical definition of a specific entropy type. For BG entropy, one just has $\Upsilon = 0$. With the assumption that all the entropies in (\ref{Aberule}) are positive, one deals with superadditivity for $\Upsilon \geq 0$, and with subadditivity for $\Upsilon <0$. In fact, the physical interpretation of $\Upsilon$  is the result of the long-range interactions between subsystems, which leads to nonadditivity. 

\subsection{Beyond extensivity}

In BG thermodynamics the additivity and the extensivity are closely related - additivity implies extensivity and extensivity implies additivity  \cite{Swendsen}. This is not the case in general, and so the extensivity and the additivity may not be related, i.e. the extensivity may not imply the additivity and vice verse. An example of such a kind of a quantity, which is based on the definition (\ref{homog}) is given by the function $f(X_1,X_2) = x_1^2/ \sqrt{X_1^2 + X_2^2}$. It obeys extensivity, but not additivity \cite{soton29487}.

Generally, the entropy $S$ is nonextensive if 
\be
S(aX) \neq a S(X) ,
\ee
where $X$ is a thermodynamical quantity and $a >0$, i.e. when the relations (\ref{homog}) and (\ref{homogS}) are violated.

\section{A comparable analysis of nonextensive entropies plethora}
\label{zoo}

\subsection{Bekenstein entropy}

The Bekenstein entropy is not motivated by anything like statistical mechanics, but it is a well established notion in gravity theory \cite{Bekenstein:1973ur}. For a Schwarzschild black hole it reads
\be
S_{Bek} = 4 \pi k_B \left( \frac{M}{m_p} \right)^2 = \frac{4 \pi k_B G M^2} {\hbar c} ,
\label{Bek}
\ee
and it is usually presented with its accompanying Hawking temperature, which reads
\be
T_H = \frac{\hbar c^3}{8 \pi Gk_B M} ,
\label{Hawk}
\ee
where $M$ is the mass of a black hole, $c$ is the speed of light, $G$ is the gravitational constant, $\hbar$ is the reduced Planck constant, and $m_p$ is the Planck mass. In fact, the temperature (\ref{Hawk}) can be calculated from the entropy (\ref{Bek}), by applying the Clausius formula 
\be
\frac{k_B}{T} = \frac{\partial S}{\partial E} ,
\label{Clausius}
\ee
and using the Einstein mass-energy equivalence formula $E = Mc^2$. 

It is not always understood in the literature that because of the area rather than volume scaling, the Bekenstein entropy is nonextensive, and that it obeys the following nonadditive composition rule (see e.g. \cite{Alonso-Serrano:2020hpb})
\be 
S_{A+B} = S_A + S_B + 2 \sqrt{S_A} \sqrt{S_B} ,
\label{Bekrule}
\ee
which we will call the {\it square root rule} from now on. This rule comes directly as a consequence of (\ref{Bek}), according to which the entropy $S_{Bek} \propto M^2$, so that $S_A \propto M_A^2$, and $S_B \propto M_B^2$. If black holes merge in an adiabatic way, then their mass after merging is the sum $M_{A+B} = M_A+M_B$, but the entropy $S_{A+B} \propto M_{A+B}^2$, giving an extra term $1/2 M_A M_B$, which is an extra  nonadditive term in (\ref{Bekrule}). 

Curiously, after making a redifinition of the Bekenstein entropies as $\tilde{S}_{A+B} \equiv \sqrt{S_{A+B}} = M_{A+B} = M_A + M_B$, $\sqrt{S_A} = M_A \equiv \tilde{S}_A$, and $\sqrt{S_B} = M_B \equiv \tilde{S}_B$, one can rewrite the composition rule (\ref{Bekrule}) in an additive way
\be
 \tilde{S}_{A+B} = \tilde{S}_A  + \tilde{S}_B ,
\ee
but this is not of any physical meaning. 

In conclusion, Bekenstein entropy addition formula (\ref{Bekrule}) does not fulfil Ab\'e rule (\ref{Aberule}), though it looks quite similar. We comment on this point in relation to other entropies in subsection \ref{Barrow}. 

\subsection{Tsallis $q$, Tsallis-Cirto $\delta$, Tsallis $q, \delta$, and Tsallis-Jensen $q, \gamma$ entropies}

\subsubsection{Tsallis $q$-entropy}

The Tsallis $q$-entropy \cite{Tsallis:1987eu,tsallisbook} is one of the earliest proposals for generalization of BG entropy. It encompasses an issue of the long-range interaction between thermodynamical subsystems by introducing a nonextensivity parameter $q$ ($ q \in R$) into the BG entropy definition (\ref{BG}) keeping the standard BG condition that the sum of all the probabilities is equal to one, i.e. that $\sum p_i = 1$. It reads as follows
 \be
\mathcal{S}_{q} = k_B \sum_{i=1}^n  p_i \ln_q \frac{1}{p_i} =  -k_B\sum_{i=1}^n \left( p_i \right)^q\ln_q p_i = - k_B \sum_{i=1}^n \ln_{2-q} p_i ,
\label{S_q}
\ee 
where a newly defined $q$-logarithmic function $\ln_q p$, is introduced:
\be
\ln_qp \equiv \frac{p^{1-q}-1}{1-q} .
\label{logq}
\ee 
In the limit $q \to 1$, one has the standard logarithm $\ln_1{p} = \ln{p}$. It is important that the $q$-logarithm does not fulfil the standard logarithm addition rule $\ln{ab} = \ln{a} + \ln{b}$, where $a, b$ are some arbitrary numbers. Instead, it fulfils a nonadditive composition rule given by 
\be
\ln_q{ab} = \ln_q{a} + \ln_q{b} + (1-q)  \ln_q{a}  \ln_q{b} ,
\label{logqrule}
\ee
which, in fact, is the origin of the Ab\'e rule (\ref{Aberule}). Interestingly, by the introduction of some specific $q-$product defined as \cite{Tsallis2009} 
\be
(x \otimes y)_q \equiv \left[ x^{1-q} + y^{1-q} - 1 \right]^{\frac{1}{1-q}} , \hspace{0.1cm} (x \ge 0, y \ge 0) ,
\label{qproduct}
\ee
one can make the rule (\ref{logqrule}) additive, i.e.
\be
\ln_q{\left[ (x \otimes y)_q \right]} = ln_q x  + \ln_q y .
\ee
It is also possible to define the $q$-exponential function 
\be
e_q^p \equiv \left[1 +(1-q) p \right]^{\frac{1}{1-q}} , 
\ee
which does not fulfil the standard exponent addition rule $e^{a+b} = e^a e^b$,  though in the limit $q \to 1$ it does, since $e_1^p = e^p$. 

Tsallis $q$-entropy definition (\ref{S_q}) is usually presented in three equivalent forms. However, using the definition of q-logarithm (\ref{logq}), all of them can be brought into the same form (cf. Appendix \ref{AppendA}):
\be
\mathcal{S}_{q} = k_B \frac{1 - \sum_{i=1}^n  \left( p_i \right)^q}{q-1} .
\label{S_q2}
\ee
It is important to mention that in order to fulfil the requirements of concavity for $S_q$ given by (\ref{concavity}), the nonextensivity parameter should be positive $q>0$ \cite{Tsallis:2012js}. 

In the limit $q \to 1$, Tsallis entropy $S_q$ given by (\ref{S_q}) or (\ref{S_q2}) reduces to BG entropy (\ref{BG}). After some check, it is possible to find that Tsallis $q$-entropy (\ref{S_q}) or (\ref{S_q2}) satisfies the  nonadditive composition Ab\'e rule (\ref{Aberule}), if one defines a nonextensivity parameter as $\Upsilon = 1-q$ (cf. Appendix \ref{AppendB}). For equal probability states (\ref{p_i_equal}), the formula (\ref{S_q2}) gives the Tsallis $q$-entropy as
\be
\mathcal{S}_{q} = k_B \ln_q n  = k_B \frac{n^{q-1} -1}{1-q} ,
\ee
which nicely shows, how it generalizes BG entropy (\ref{BGequalp}) via a new parameter $q$. 

\subsubsection{R\'enyi entropy}

The R\'enyi entropy \cite{Renyi1}, which is in fact a measure of entanglement in quantum information theory, is additive and preserves event independence. It is another important generalization of BG  entropy, which is defined by
\be
S_R= k_B \frac{\ln \sum_{i=1}^n \left( p_i \right)^q}{1-q}. 
\label{SR}
\ee
By assuming that all the states are equally probable, as in (\ref{p_i_equal}), it follows from (\ref{SR}) that 
\be
S_R = k_B \ln{n} ,
\label{SReq}
\ee
which is the same as BG entropy (\ref{BG}).  

In fact, the R\'enyi entropy (\ref{SR}) can be written in terms of the Tsallis $q$-entropy by using the formal logarithm approach \cite{PhysRevE.83.061147} as follows
\begin{eqnarray}
S_R=\frac{k_B}{1-q}\ln[1+ \frac{1-q}{k_B} \mathcal{S}_q] .
\label{srenyi}
\end{eqnarray}
Quite a unique feature of R\'enyi entropy is that {\it it is additive}, which results from some more general Ab\'e composition rule given by \cite{Alonso-Serrano:2020hpb}
\be
H(S_{A+B}) = H(S_A) + H(S_B) + \frac{\Upsilon}{k_B}  H(S_A) H(S_B) ,
\ee
together with redefinition using the logarithm in the form
\be
L(S) = \frac{k_B}{\Upsilon} \ln{\left( 1 + \frac{\Upsilon}{k_B} H(S) \right)}  ,
\ee
which applied to (\ref{srenyi}) gives an additive formula
\be
L(S_{A+B}) = L(S_A) + L(S_B) ,
\label{Rcomposition}
\ee
where $L(S)$ corresponds to the R\'enyi entropy, and $H(S)$ corresponds to the Tsallis $q$-entropy. In such a formulation, one can write that the R\'enyi entropy fulfils the Ab\'e rule (\ref{Aberule}),  
with the parameter $\Upsilon = 0$.

\subsubsection{Tsallis-Cirto $\delta$-entropy}

The Tsallis-Cirto $\delta$-entropy \cite{Tsallis:2012js,Tsallis:2019giw}, sometimes also known in the literature as just the Tsallis entropy, is yet another generalization of BG entropy (\ref{BG}) by the introduction 
of another nonextensivity parameter  $\delta$ as follows 
\be
\mathcal{S}_{\delta} = k_B\sum_{i=1}^n p_i \left( \ln p_i \right)^{\delta} \hspace{0.3cm} (\delta > 0, \delta \in R) ,
 \label{Sdelta}
\ee
and this difference is easily recognized, when one compares it with the Tsallis $q$-entropy (\ref{S_q}), and with BG entropy (\ref{BG}). 

Under the assumption that all the states are equally probable, as in (\ref{p_i_equal}), one gets from (\ref{Sdelta}) that
\be
 \mathcal{S}_{\delta} = k_B \left(\ln n \right)^{\delta} \equiv k_B \ln^{\delta} n .
 \label{TCequal}
 \ee
Making another assumption that we deal with two independent systems $A$ and $B$, fulfilling the condition $n^{A+B} = n^A \cdot n^B$, one realizes that the composition rule for the Tsallis-Cirto entropy (\ref{TCequal}) reads
\begin{equation}
    \left( \frac{S_{\delta, A+B}}{k_B} \right)^{1/\delta} = \left (\frac{S_{\delta, A}}{k_B}\right)^{1/\delta}+\left (\frac{S_{\delta, B}}{k_B}\right)^{1/\delta}  ,
    \label{tsallisdeltacomp}
\end{equation}
which is another example of a composition rule, which is {\it different} from the Ab\'e rule (\ref{Aberule}). We start calling it {\it $\delta$-addition rule} from now on. In fact, Tsallis and Cirto suggest that \cite{Tsallis:2012js,Tsallis:2019giw}
\begin{eqnarray}
S_{\delta}  = k_B \left(\frac{S_{Bek}}{k_B} \right)^\delta ,
\label{deltas}
\end{eqnarray}
where $S_{Bek}$ is the Bekenstein entropy (\ref{Bek}). According to a new composition rule (\ref{tsallisdeltacomp}), one realizes that the Bekenstein entropy, as given by $S_{Bek} \propto (S_{\delta})^{1/\delta}$, can be additive, while the Tsallis-Cirto entropy $S_{\delta}$ itself is nonadditive. Besides, bearing in mind the definition of Bekenstein entropy for a Schwarzschild black hole (\ref{Bek}), one can easily notice that for $\delta=3/2$ the Tsallis-Cirto entropy (\ref{deltas}) is proportional to the volume $S_{\delta} \propto M^3$, and so it is an extensive quantity in view of the standard definition of extensivity. We come back to this issue in subsection \ref{Barrow}.

\subsubsection{Tsallis $q, \delta$-entropy}

The Tsallis $q,\delta$-entropy generalizes both the Tsallis $q$-entropy (\ref{S_q}) and the Tsallis-Cirto $\delta$-entropy (\ref{Sdelta}), combining them as follows \cite{Tsallis:2012js,Tsallis:2019giw} 
\be
\mathcal{S}_{q, \delta} = k_B\sum_{i=1}^n p_i \left( \ln_q p_i \right)^{\delta} \hspace{0.3cm} (\delta > 0, q \in R, \delta \in R) .
 \label{Sqdelta}
\ee
Now both $q$ and $\delta$ play the role of two independent nonextensivity parameters.  
By assuming that all the states are equally probable as in (\ref{p_i_equal}), one gets from (\ref{Sqdelta}) that
\be
 \mathcal{S}_{q, \delta} = k_B \left(\ln_q n \right)^{\delta} \equiv k_B \ln_q^{\delta} n .
 \ee
The Tsallis $q, \delta$-entropy fulfils neither Ab\'e addition rule nor $\delta$-addition rule, though it does the former in the limit $\delta \to 0$, and the latter in the limit $q \to 0$. 

\subsubsection{Tsallis-Jensen $q, \gamma$-entropy}
\label{TJEntropy}

Recently, Tsallis and Jensen \cite{Tsallis2024} proposed another generalization of BG entropy which reads 
\be
S_{q, \gamma} = k_B \left[ \frac{ \ln \sum_{i=1}^{n} p_i^q} {1-q} \right]^\frac{1}{\gamma} =  k_B \left( \frac{S_R}{k_B} \right)^{\frac{1}{\gamma}} ,
\label{STJ}
\ee
where $S_R$ is the R\'enyi entropy and $\gamma$ is a new parameter somewhat analogous to the parameter $\delta$ in the Tsallis-Cirto entropy (\ref{Sdelta}).

Since the R\'enyi entropy possesses BG limit for $q \to 1$, then we can write \cite{Tsallis2024} 
\be
S_{1,\gamma} = k_B \left( \frac{S_{BG}}{k_B}  \right)^{\frac{1}{\gamma}}  ,
\label{STJ_1g}
\ee
and analogously, if we take Bekenstein entropy (\ref{Bek}) instead of the BG in (\ref{STJ_1g}), in the same limit $q \to 1$, we get
\be
 S_{1,\gamma}^{Bek} = k_B \left( \frac{S_{Bek}}{k_B} \right)^{\frac{1}{\gamma}}  .
\label{SBek_1g}
\ee
Bearing in mind the additive composition rule (\ref{addSrule}), for the BG entropy, and using (\ref{STJ_1g}), one can write the additivity rule for $S_{1,\gamma}$ in such a case as
\be
\left[ S_{1,\gamma}(A+B) \right]^{\gamma} = \left[ S_{1,\gamma}(A) \right]^{\gamma} +  \left[ S_{1,\gamma}(B) \right]^{\gamma} .
\ee
Similarly, taking into account the square root additivity rule (\ref{Bekrule}) for the Bekenstein-like entropy (\ref{SBek_1g}), one can write the composition rule as 
\be
\left[ S_{1,\gamma}^{Bek}(A+B) \right]^{\gamma} = \left[ S_{1,\gamma}^{Bek}(A) \right]^{\gamma} +  \left[ S_{1,\gamma}^{Bek}(B) \right]^{\gamma} + 
2  \left[ S_{1,\gamma}^{Bek}(A)  S_{1,\gamma}^{Bek}(B) \right]^{\frac{\gamma}{2}}.
\ee
Finally, since the R\'enyi entropy $S_R$ in (\ref{STJ}) is in general additive according to the composition rule (\ref{Rcomposition}), then we can 
write a generic composition rule for the Tsallis-Jensen entropy (\ref{STJ}) as
\be
 \left[ S_{q,\gamma}(A+B) \right]^{\gamma} = \left[ S_{q,\gamma}(A) \right]^{\gamma} +  \left[ S_{q,\gamma}(B) \right]^{\gamma} ,
\label{STJ_compo}
\ee
and this is exactly the $\delta$ composition rule (\ref{tsallisdeltacomp}) with $\gamma = 1/\delta$.

\begin{table*}
\caption{Tsallis entropies}
\begin{center}
\label{TsallisEntropies}
\begin{tabular}{lccccc}
\hline
\\
Entropy Type & Extensivity & Additivity & Ab\'e addition rule & $\delta-$addition rule\\
\hline
\\
Boltzmann-Gibbs $S_{BG}$ & yes & yes & yes, $\Upsilon = 0$& yes, $\delta =1$\\
Tsallis $S_{q,1}=S_q$ & no & no & yes, $\Upsilon = 1-q$  & no \\
Tsallis-Cirto $S_{1,\delta}=S_{\delta}$ & no & no & no & yes \\
General Tsallis $S_{q,\delta}$   & no & no &  no & no \\
Tsallis-Jensen $S_{q,\gamma}$ & no & no & no & yes, $\delta = 1/\gamma$ \\
\\
\hline
\end{tabular}
\end{center}
\end{table*}

Table \ref{TsallisEntropies} gives the summary of four different Tsallis invented entropies. 

\subsection{Barrow fractal horizon $\Delta$-entropy and its relation to Bekenstein and Tsallis-Cirto $\delta$-entropy}
\label{Barrow}

Barrow entropy \cite{Barrow:2020tzx} has no statistical roots at all. It is closely tied to black hole horizon geometry influenced by quantum fluctuations, which make initially smooth black hole horizon a fractal composed of spheres, forming the so-called sphereflake. This structure is characterised by the fractal dimension $d_f$ which in the extreme cases are either the surface or the volume i.e. $2 \leq d_f \leq 3$, and results in an effective horizon area of $r^{d_f}$, where $r$ is the black hole horizon radius. After quantum-motivated modification of the area, the entropy reads   
\begin{eqnarray}
S_{Bar}=k_B \left( \frac{A}{A_p}\right)^{1+\frac{\Delta}{2}} = k_B \left( \frac{S_{Bek}}{k_B}\right)^{1+\frac{\Delta}{2}} ,
\label{barrowarea}
\end{eqnarray}
where $S_{Bek}$ is Bekenstein entropy, $A$ - the horizon area, $A_p$ - the Planck area, $A_p \propto l_p^2$  with $l_p$ - Planck length, and $\Delta$ is the parameter related to the fractal dimension by the relation $\Delta=d_f-2$. In fact, $0 \leq \Delta \leq 1$ with $\Delta\rightarrow1$ limit yielding maximally fractal structure, where the horizon area behaves effectively like a $3-$dimensional volume,  and with $\Delta \rightarrow 0$ limit yielding the Bekenstein area law, where no fractalization occurs. Although Barrow entropy has geometrical roots, and is not motivated by thermodynamics, it has the same form as Tsallis-Cirto $\delta$ entropy (\ref{deltas}) \cite{Abreu:2020wbz} being also related to Bekenstein entropy $S_{Bek}$, as in (\ref{Bek}), provided that 
\begin{eqnarray}
\delta = 1+\frac{\Delta}{2} .
\label{barrowdelta}
\end{eqnarray}
However, the ranges of parameters $\delta$ and $\Delta$ are different - $\delta$ has only the bound $\delta >0$, while $0 \leq \Delta \leq 1$ is equivalent to $1 \leq \delta \leq 3/2$. 
Thus, qualitatively, both entropic forms yield the same temperatures as a function of a black hole mass. Both Tsallis-Cirto entropy limit $\delta \to 3/2$ and Barrow limit $\Delta \to 1$ yield an extensive, but still nonadditive entropy for black holes. In fact, the cosmological studies of Barrow entropy as holographic dark energy have been performed intensively \cite{Saridakis:2020zol,MPD_Salz_2020,Saridakis:2020lrg,AlMamon:2020usb,Asghari:2021lzu,Adhikary:2021xym,Nojiri:2021czz,Drepanou:2021jiv,DiGennaro:2022ykp,DiGennaro:2022grw,Nojiri:2021jxf,Anagnostopoulos:2020ctz,Sheykhi:2018dpn,FeiziMangoudehi:2022rwj,Ghoshal:2021ief,Luciano2022a,Leon:2021wyx,Almada2022_Kaniad_test}, pointing towards this extensive case \cite{MPD_Salz_2020,PhysRevD.108.103533,Tsallis2024}.

\subsection{Landsberg $U$-entropy}

The Landsberg $U$-entropy is defined in relation to Tsallis $q-$entropy (\ref{S_q2}) as  \cite{soton29487}
\be
S_U = \frac{k_B}{1-q} \left( 1 - \frac{1}{\sum_{i=1}^{n} \left( p_i \right)^q} \right) =  k_B \frac{1 - \sum_{i=1}^n  \left( p_i \right)^q}{q-1}  \frac{1}{\left( p_i \right)^q} 
= \frac{\mathcal{S}_{q}}{\sum_{i=1}^{n} \left( p_i \right)^q} ,
\label{SU}
\ee
and it fulfils the Ab\'e rule (\ref{Aberule}) for $\Upsilon = q-1$ (cf. Appendix \ref{AppendB}). By assuming that all the states are equally probable as in (\ref{p_i_equal}), it simplifies (\ref{SU}) to the form
\be
S_U = n^{q-1} S_q ,
\label{rel_Uq}
\ee
so it simply relates to Tsallis $q$-entropy. 



\subsection{Sharma-Mittal entropy}

The Sharma-Mittal (SM) entropy \cite{SM,MASI2005217} combines the R\'enyi entropy with the Tsallis $q-$entropy, and is defined as
\begin{eqnarray}
    S_{SM}= \frac{k_B}{R} \left[\left(\sum_{i=1}^n (p_i)^{q}\right)^{\frac{R}{1-q}}-1 \right]  ,
    \label{SM}
\end{eqnarray}
where $R$ is another dimensionless parameter apart from $q$.  For equally probable states in (\ref{p_i_equal}), one gets from (\ref{SM}) that \cite{SayahianJahromi:2018irq}
\begin{eqnarray}
S_{SM}=\frac{k_B}{R}\left\{\left[1+\frac{1-q}{k_B} S_{q} \right]^{\frac{R}{1-q}} -1\right\} ,
\label{SSM}
\end{eqnarray}
where $R \rightarrow 1-q$ limit yields the Tsallis entropy, and $R\rightarrow 0$ limit yields the R\'enyi entropy. It is interesting to notice that the SM entropy obeys the composition rule of Ab\'e (\ref{Aberule}) for $\Upsilon = 1$ (cf. Appendix \ref{AppendC}).

\subsection{Kaniadakis entropy}

The Kaniadakis entropy \cite{Kaniadakis:2002zz,Kaniadakis:2005zk,Drepanou:2021jiv} results from taking into account Lorentz transformations of special relativity. It is a single $K$-parameter ($-1<K<1$) deformation of BG entropy (\ref{BG}), with $K$ parameter related to the dimensionless rest energy of the various parts of a multibody relativistic system. The basic definition of Kaniadakis entropy, which directly generalizes BG entropy reads 
\be
S_K = - k_B \sum_{i=1}^n p_i \ln_K {p_i} ,
\label{SKln}
\ee
where $p_i$ is the probability distribution and $n$ is the total number of states, as mentioned in Section \ref{BGT}. The formula (\ref{SKln}) introduces the $K$-logarithm 
\be
\ln_K {x} \equiv \frac{x^K - x^{-K}}{2K}  = \frac{1}{K} \sinh{(K \ln x)} 
\label{lnK}
\ee
with some basic properties like $\ln_K x^{-1} = - \ln_K x$ and $\ln_{-K} x = \ln_K x$, giving the standard logarithm $\ln x$ in the limit $K \to 0$. An equivalent definition of Kaniadakis entropy, which can be obtained after the application of $K-$logarithm (\ref{lnK}) reads 
\be
S_K = - k_B \sum_{i=1}^n \frac{(p_i)^{1+K} - (p_i)^{1-K}}{2K} .
\label{SK}
\ee
The $K-$deformed logarithm is associated with the $K-$exponential as follows
\be
{\rm exp}_K x = \exp{\left[\frac{1}{K} {\rm arcsinh} (Kx) \right]} = \left(\sqrt{1 +K^2 x^2} +Kx \right)^{1/K}  ,
\label{expK}
\ee
and it fulfils some basic relations like ${\rm exp}_K (x) {\rm exp}_K (-x) = 1$, and ${\rm exp}_K (x) {\rm exp}_{-K} (x)$, giving the standard exponential function $\exp{(x)}$ in the limit $K \to 0$. In fact, $K-$logarithm and $K-$exponential are the inverse functions, which means that they fulfil the relation
\be
\ln_K{\left({\rm exp}_K x \right)} = {\rm exp}_K \left( \ln_K x \right) = x  .
\ee
The $K-$logarithm fulfils a generalized composition rule which reads 
\be
\ln_K (xy) = \ln_K x \sqrt{1 + K^2 (\ln_K y)^2} + \ln_K y \sqrt{1 + K^2 (\ln_K x)^2}  , 
\label{Kxy}
\ee
and it admits the standard logarithm rule $\ln(xy) = \ln x + \ln y$, in the limit $K \to 0$. The rule (\ref{Kxy}) comes from the definition of $K-$sum 
\be
(x \oplus y)_K = x \sqrt{1+K^2y^2} + y \sqrt{1+K^2x^2} , 
\label{Ksum}
\ee
where one replaced $x \to \ln x$ and $y \to \ln y$, and gives standard additivity rule $(x \oplus y)_K = x + y$ in the limit $K \to 0$. Using the definition of Kaniadakis entropy (\ref{SKln}), as well as the $K-$logaritm addition rule, we can write down the Kaniadakis entropy composition rule as follows 
\be
S_K(A+B) = S_K(A) \sqrt{1 + \frac{K^2}{k_B^2} S_K(B)} + S_K(B) \sqrt{1 + \frac{K^2}{k_B^2} S_K(A)}  ,
\ee
which we start calling {\it $K-$addition rule} from now on. 
It is interesting to note that by the application of the $K-$sum \cite{Kaniadakis:2002zz} defined as 
\be
(x \otimes y)_K = \frac{1}{K} \sinh{ \left[ \frac{1}{K} {\rm arcsinh} (Kx) {\rm arcsinh}  (Ky) \right] } ,
\label{Kproduct}
\ee
one has for the $K-$logarithm
\be
\ln_K \left[ (x \otimes y)_K \right] = \ln_K x + \ln_K y , 
\ee
so that applying it to (\ref{SKln}), the Kaniadakis entropy (in full analogy to the $q-$product of Tsallis given by (\ref{qproduct})), can take a completely {\it additive form} as below (cf. the definition of additivity (\ref{addrule}) for statistically independent systems) 
\be
S_K (A+B)_K = S_K(p_A p_B) = S_K (p_A) + S_K (p_B) = S_K (A) + S_K (B) . 
\ee
Using the $K-$product, the Kaniadakis entropy can also be {\it extensive} 
\be
S_K (x^{\otimes r}) =  r S_K (x)  , 
\ee
 where $r=$ const.,  and it comes as a result of the identity 
 \be
 \ln_K (x^{\otimes r}) = r \ln_{rK} (x) .
 \ee
Finally, in analogy to the previous considerations, and under the assumption that all the states are equally probable, as in (\ref{p_i_equal}), one gets from (\ref{SKln}) that 
\be
\ln_K{p_i} = - \frac{1}{K} \frac{e^{K \ln{n}} - e^{-K \ln{n}}}{2} ,
\ee
which can further be transformed into 
\be
S_K = \frac{k_B}{K} \sinh{\left( \frac{K}{k_B} S \right)} ,
\label{SKsh}
\ee
where $S = k_B \ln{n}$ is the BG entropy (\ref{BGequalp}).


\subsection{Thermal equilibrium temperature vs equilibrium temperature for nonextensive entropies}

One of the important issues related to the nonextensive systems is the formulation of a proper definition of the thermal equilibrium temperature which is necessary according to the zeroth law of thermodynamics \cite{PhysRevE.67.036114,OU2006525,PhysRevE.83.061147}. This problem can be solved by using the notion of the {\it effective equilibrium temperature} based on the equilibrium condition \cite{Abe_2001_PLA}, recently discussed in more detail in Ref. \cite{Cimdiker:2022ics}. In defining this temperature, one uses an analogy to some strongly coupled quantum systems which can be in equilibrium at the effective temperature, but not in the thermal equilibrium, as for the extensive systems. The equilibrium temperature is obtained by maximising the composition rule (\ref{Composable}) with the fixed total energy of the system $U_{AB} = U_A + U_B$ which leads to the conditions $\delta S_{AB} = 0$, with $S_{AB} = g(S_A,S_B)$, and $\delta U(A+B)=0$. The equilibrium temperature $T_{eq}$ then reads
\be
T_{eq} \equiv \frac{1}{k_B \beta^{*}} = \frac{\frac{\partial g}{\partial S_B}}{k_B \beta_A} =  \frac{\frac{\partial g}{\partial S_A}}{k_B \beta_B},
\ee
where we have defined $\beta$ for each system i.e.
\be
k_B \beta_A =  \frac{\partial S_A}{\partial U_A} \hspace{0.2cm} {\rm and}  \hspace{0.2cm} k_B \beta_B =  \frac{\partial S_B}{\partial U_B} .
\ee
The application of the above procedure  to the specific system described by the Tsallis $q$-entropy (\ref{S_q}), which fulfils the Ab\'e composition rule (\ref{Aberule}) as an example of the general composition rule (\ref{Composable}), gives the equilibrium temperature  \cite{Cimdiker:2022ics}
\be
T_{eq} = T_{R}=(1+\frac{1-q}{k_B} S_{q})\frac{1}{k_B \beta} ,
\label{T_eqR}
\ee
which is the R\'enyi temperature $T_R$ \cite{Cimdiker_2023} corresponding to R\'enyi entropy $S_R$ given by (\ref{srenyi}) according to the Clausius formula (\ref{Clausius}). 

It is interesting that the R\'enyi entropy and the R\'enyi effective equilibrium temperature, can be defined on a horizon of a Schwarzschild black hole \cite{Biro:2013cra,Czinner:2015ena,Czinner:2015eyk,Czinner:2017tjq,Czinner:2017bwc}, by assuming that the Bekenstein entropy $S_{Bek}$ given by (\ref{Bek}) is the Tsallis entropy $S_q$ in (\ref{srenyi}) and (\ref{T_eqR}).

Similar procedure of introducing the equilibrium temperature can be performed for the Tsallis-Cirto $\delta$-entropy by calculating the corresponding temperature from the Clausius relation (\ref{Clausius}) as follows \cite{Cimdiker:2022ics,Cimdiker_2023}
\be
T_\delta=\frac{T_{H}}{\delta}\left(\frac{S_{Bek}}{k_B}\right)^{1-\delta} , \label{Tdelta} 
\ee 
which scales with $1/M^2$ for $\delta=3/2$, i.e., $T_\delta  \propto 1/M^2$ for a Schwarzschild black hole.

\subsection{Classification of entropies}

Bearing in mind all the considerations of the whole Section \ref{zoo}, we present the summary of the additivity and extensivity properties of entropies in the Table \ref{addext_table}. 

There is the whole group of Tsallis invented thermodynamical entropies, which generalize BG entropy in some different ways (cf. also Table \ref{TsallisEntropies}). They obey either the Ab\'e composition rule or $\delta$-addition rule. The Tsallis $q$-entropy relates to both the R\'enyi and the Landsberg ${\cal U}$ entropies, while it is generalized by the Sharma-Mittal entropy. On the other hand, the Tsallis-Cirto $\delta$-entropy is related to the Barrow entropy and the Tsallis-Jensen $q,\gamma$ entropy and, interestingly, it points towards extensivity, when observations are taken into account  \cite{MPD_Salz_2020,PhysRevD.108.103533,Tsallis2024}. The Kaniadakis entropy seem to form a separate group of nonextensive entropies because of its hyperbolic formulation as a consequence of relativity theory being taken into account, but it stil has a BG limit. In fact, all the entropies in our study have BG limit, except Bekenstein entropy, which is not formulated as proper statistics. However, it is composable though its composition rule is unique among any other rules, which are often shared with themselves. Besides, in the microscopic counting of states in the string theory \cite{strominger1996,DEHARO2020}, the nonextensive Bekenstein formula is recovered, though it was also found that some higher-dimensional extremal black holes allow Bekenstein entropy, which is  consistent with the Boltzmann-Gibbs extensive limit. 

\begin{table*}
\caption{The additivity, extensivity, and composability properties of entropies.}
\label{addext_table}
\begin{center}
\begin{tabular}{lccccc}
\hline
\\
Entropy Type & Extensivity & Additivity & Ab\'e rule & $\delta-$rule & $K-$rule\\
\hline
\\
Boltzmann-Gibbs  $S_{BG}$ & yes &  yes & yes,  $\Upsilon =0$ & yes, $\delta=1$ & yes, $K=0$\\
Bekenstein $S_{Bek}$ & no & no* & no & no & no  \\
Tsallis $q$-entropy $S_q$ & no & no &  yes, $\Upsilon = 1-q$ & no & no \\
Tsallis-Cirto $S_{\delta}$ $(\delta \neq \frac{3}{2})$ & no & no & no & yes & no\\
Tsallis-Cirto $S_{\delta}$ $(\delta = \frac{3}{2})$ & yes & no & no & yes, $\delta= \frac{3}{2}$ & no\\
Barrow $S_{Bar} = S_{Bek}$ $(\Delta = 0)$ & no & no* & no & no & no \\
Barrow $S_{Bar}$ $(0 < \Delta <1)$ & no & no & no & yes & no \\
Barrow $S_{Bar}$ ($\Delta =1)$ & yes & no & no & yes,  $\delta= \frac{3}{2}$ &  no\\
R\'enyi $S_R$ & no & yes & yes, $\Upsilon = 0$ & no & no \\
Landsberg $U$-entropy $S_U$ & no  & no & yes, $\Upsilon = q-1$ & no & no \\
Kaniadakis $S_K$ & no & no  & no & no & yes\\
Sharma-Mittal $S_{SM}(q,R)$  & no & no  & yes, $\Upsilon = R$ & no & no \\
Tsallis $q,\delta$-entropy $S_{q,\delta}$ & no & no & no & no & no \\
Tsallis-Jensen $S_{q,\gamma}$ & no & no & no & yes, $\delta = 1/\gamma$ & no \\
Tsallis-Jensen $S_{1,\gamma}$ & no & no & no & yes, $\delta = 1/\gamma$ & no\\
Tsallis-Jensen $S_{q,1}=S_R$ & no & yes & yes, $\Upsilon = 0$ & no & no \\
\\
\hline
* obeys square root composition rule (\ref{Bekrule})
\end{tabular}
\end{center}
\end{table*}

\subsection{Generalized four- and five-parameter entropic forms}

There exists a four-parameter entropic formula \cite{Nojiri:2022dkr}, which reads 
\be
S_g(\alpha_{\pm},\beta,\sigma) = \frac{k_B}{\sigma} \left[\left(1 + \frac{\alpha_+}{\beta} \frac{S}{k_B} \right)^{\beta} - \left(1 + \frac{\alpha_-}{\beta} \frac{S}{k_B} \right)^{-\beta} \right] ,
\ee
as well as the five-parameter formula \cite{Odintsov:2022qnn} of the form 
\be
S_g(\alpha_{\pm},\beta,\sigma,\epsilon) = \frac{k_B}{\sigma} \left\{ \left[ 1 + \frac{1}{\epsilon} \tanh{\left( \frac{\epsilon \alpha_+}{\beta} \frac{S}{k_B} \right)} \right]^{\beta} -  \left[ 1 + \frac{1}{\epsilon} \tanh{\left( \frac{\epsilon \alpha_-}{\beta} \frac{S}{k_B} \right)} \right]^{-\beta}  \right\},
\ee
Both these formulas generalize some of the entropies which are contained in the Table \ref{addext_table} and have the following limits:
\begin{enumerate} 
\item if $\epsilon \to 0$, then one recovers Tsallis-Cirto (\ref{Sdelta}) and Barrow (\ref{barrowarea}) entropies; 
\item if $\epsilon \to 0$, $\alpha_- \to 0$, $\beta \to 0$, and $\alpha_+/\beta$ finite, then one recovers R\'enyi entropy (\ref{srenyi});
\item if $\epsilon \to 0$, $\alpha_- \to 0$, $\sigma = \alpha_+ = R$, and $\beta = R/\delta$, then one recovers Sharma-Mittal entropy formula (\ref{SSM}), though only when one replaces Tsallis $q$-entropy $S_q$ with Tsallis-Cirto $\delta$-entropy $S_{\delta}$; 
\item if $\epsilon \to 0$, $\beta \to \infty$, $\alpha_+ = \alpha_- = \sigma/2 = K$, then one recovers Kaniadakis entropy (\ref{SKsh}).
\end{enumerate} 

These entropies have an important advantage for cosmology. Namely, they are singular-free at the cosmological bounce, which is characterized by vanishing of the Hubble parameter $H=0$ in a bouncing scenarios \cite{Nojiri:2022aof}. Besides, they allow microscopic interpretation \cite{Nojiri:2023bom,Nojiri:2023ikl}. 

Since all of them are in general nonadditive and nonextensive, then it is not necessary to discuss them in more detail, though they may recover BG behaviour in some special cases. However, we will not consider these cases in detail here. 

\section{Summary and discussion}
\label{summary}

Beginning with the underlying properties of the Boltzmann-Gibbs classical entropy, we have investigated the problem of nonextensity and nonadditivity in thermodynamics aiming towards gravitational systems which admit long-range interactions. The focus has been on such extensions of Boltzmann-Gibbs entropic form, which allow various deformations of it, via some new parameters modifying the space of microstates $\Omega$. These new parameters are given some interpretations according to a deformation and can be enumerated as: Tsallis nonextensivity $q-$parameter, Tsallis-Cirto nonextensivity $\delta-$parameter which is equivalent to Barrow fractality $\Delta-$parameter, Tsallis-Jensen nonextensivity parameter $\gamma$, Sharma-Mittal $R-$parameter, and Kaniadakis relativistic $K-$deformation parameter. 

The entropies under study may fulfil some composition rules which are the Ab\'e rule, and $\delta-$addition rule, both of which are nonaddtitive. The Bekenstein entropy obeys some other nonadditive rule, called the square root rule, which is somewhat similar, but not the type of the Ab\'e rule. The Kaniadakis entropy can be made additive within a special $K-$deformed algebra, and it reaches the standard Boltzmann-Gibbs additive rule in the limit $K \to 0$. The same may refer to other entropies which are composable, such as the Tsallis $q$-entropy and the Bekenstein entropy. A separate matter is whether these "made-additive" entropies have a proper physical interpretation. Another point that we have discussed was the definition of the equilibrium temperature for nonextensive systems in contrast to the thermal equilibrium temperature, which is ambiguous for these systems.

We have presented two important comparative tables with the additivity, extensivity, and composability properties of the entropies under study, and the relations between them have been shown explicitly. Up to our knowledge, such a collective comparison of entropic forms have not yet been presented in the literature. In view of recent interest of both astrophysicists and cosmologists in the application of the plethora of alternative to Boltzmann-Gibbs entropies, this paper may then serve a useful guide to these applications. It is worth noting that a number of these entropies are considered suitable as the candidates for the holographic dark energy and so the presented summary gives some structure to their general properties in a systematic way, which can help astrophysicists and cosmologists sort them out into categories which can then be systematically verified by the data. Nowadays, it seems that most of the tests are being done mainly for some particular types of entropies without much regard to the relation to others.

\vspace{6pt} 




\funding{This research was partially supported by the Polish National Science Centre grant No. DEC-2020/39/O/ST2/02323.} 

\acknowledgments{The discussions with Ilim \c{C}imdiker and Hussain Gohar are acknowledged. }

\conflictsofinterest{The author declares no conflict of interest.}



\appendixtitles{no} 
\appendixsections{multiple} 
\appendix
\section{Equivalent forms of Tsallis $q$-entropy}
\label{AppendA}

There are three equivalent forms of Tsallis $q-$entropy given by (\ref{S_q}), which we can label according to their appearance in (\ref{S_q}): $S_{qI}, S_{qII}, S_{qIII}$. We show their equivalence, by reducing each of them to the form (\ref{S_q2}) by the application the probability sum rule $\sum p_i = 1$ and the formula (\ref{logq}). For $S_{qI}$ one has
 \bea
\mathcal{S}_{qI} \equiv k_B \sum_{i=1}^n  p_i \ln_q \frac{1}{p_i} = k_B \sum_{i=1}^n p_i \frac{\left( \frac{1}{p_i} \right)^{1-q} - 1}{1-q} = k_B \sum_{i=1}^n \frac{(p_i)^q - 1}{1-q} = k_B  \frac{1 - \sum_{i=1}^n (p_i)^q}{q-1} , 
\label{S_I}
\eea
which is equivalent to (\ref{S_q2}). For $S_{qII}$, one has 
\bea
\mathcal{S}_{qII} =  - k_B\sum_{i=1}^n \left( p_i \right)^q\ln_q p_i = - k_B \sum_{i=1}^n (p_i)^q  \frac{(p_i)^{1-q} - 1}{1-q} = - k_B \frac{ \sum_{i=1}^n  \left[ (p_i)  - (p_i)^q \right]}{1-q} =  k_B  \frac{1 - \sum_{i=1}^n (p_i)^q}{q-1} ,
 \label{S_II} 
\eea
which is equivalent to (\ref{S_q2}). For $S_{qIII}$ one has 
\bea
\mathcal{S}_{qIII} = - k_B \sum_{i=1}^n \ln_{2-q} p_i = - k_B \sum_{i=1}^n (p_i)  \frac{(p_i)^{q-1} - 1}{q-1} =  - k_B \sum_{i=1}^n \frac{(p_i)^q - (p_i) }{q-1} = k_B  \frac{1 - \sum_{i=1}^n (p_i)^q}{q-1} , 
\label{S_III} 
\eea
 where we have applied a redefined $q-$logarithm (\ref{logq}) with $q' = 2-q$. This finishes the prove.

\section{Validity of Ab\'e composition rule for Tsallis $q-$entropy and Landsberg $U-$entropy} 
\label{AppendB}

We first assume two probabilistically independent systems $A$ and $B$ fulfilling 
\be
 \sum_{i=1}^n \sum_{j=1}^m p_{ij}^{A+B} =  \sum_{i=1}^n \sum_{j=1}^m p_i^A p_j^B 
\label{multi}
\ee
for every $i, j$ which gives 
\bea
 \sum_{i=1}^n \sum_{j=1}^m \left( p_{ij}^{A+B} \right)^q  =   \sum_{i=1}^n \sum_{j=1}^m \left( p_i^A p_j^B \right)^q =  \sum_{i=1}^n  \sum_{j=1}^m \left( p_i^A\right)^q \left( p_j^B \right)^q .
 \label{multi_sum}
\eea
Following the definition of Tsallis $q-$entropy  (\ref{S_q2}), one can write using (\ref{multi_sum}):
\bea
S_q(A+B) = k_B  \frac{1 - \sum_{i=1}^n \sum_{j=1}^m \left( p_{ij}^{A+B} \right)^q}{q-1} = k_B  \frac{1 - \sum_{i=1}^n  \sum_{j=1}^m \left( p_i^A \right)^q \left( p_j^B \right)^q}{q-1} ,
\eea
as well as 
\be
S_q(A) = k_B  \frac{1 - \sum_{i=1}^n  \left( p_i^A \right)^q }{q-1} ,
\ee
and analogously for $S_q(B)$. 

Following Ab\'e rule (\ref{Aberule}) we can then write the right-hand side of it as 
\bea
&& RHS = S_q(A) + S_q(B) + \frac{\Upsilon}{k_B} S_q(A) S_q(B) = k_B  \frac{1 - \sum_{i=1}^n  \left( p_i^A \right)^q }{q-1}  
+ k_B  \frac{1 - \sum_{j=1}^m  \left( p_j^A \right)^q }{q-1}  \nonumber  \\ 
&& +  \Upsilon k_B  \left( \frac{1 - \sum_{i=1}^n  \left( p_i^A \right)^q }{q-1}\right)   
 \left( \frac{1 - \sum_{j=1}^m  \left( p_i^A \right)^q }{q-1} \right) = \frac{k_B}{q-1} \left[ 1 - \sum_{i=1}^n (p_i^A)^q + 1 - \sum_{i=1}^n (p_j^B)^q \right. \nonumber \\
 && \left. + \frac{Y}{q-1} - \frac{Y}{q-1}   \sum_{i=1}^n (p_i^A)^q - \frac{Y}{q-1}   \sum_{i=1}^n (p_j^B)^q + \frac{Y}{q-1}  \sum_{i=1}^n (p_i^A)^q  \sum_{j=1}^m (p_j^B)^q \right]  ,
\eea
which after selecting $\Upsilon = 1-q$, cancels six out of eight terms giving on the base of (\ref{multi_sum}) that
\bea
RHS = k_B \frac{1 - \sum_{i=1}^n \sum_{j=1}^m \left( p_i^A\right)^q \left( p_j^B \right)^q }{q -1} =  k_B \frac{1 - \sum_{i=1}^n \sum_{j=1}^m \left( p_{ij}^{A+B} \right)^q }{q -1} = S_q(A+B) .
\eea

The proof of applicability of the Ab\'e rule for Landsberg $U-$entropy (\ref{SU}), proceeds analogously, provided that $\Upsilon=q-1$, instead. 

\section{Validity of Ab\'e composition rule for Sharma-Mittal entropy} 
\label{AppendC}

Let us write the Ab\'e rule for the Sharma-Mittal entropy as follows
\be
S_{SM}(A+B) = S_{SM}(A) + S_{SM}(B) + \frac{\Upsilon}{k_B} S_{SM}(A)  S_{SM}(B)
\label{SMrule}
\ee
with 
\be
 S_{SM}(A+B) = \frac{k_B}{R} \left[\left(\sum_{i=1}^n \sum_{j=1}^m (p_{ij}^{A+B})^{q}\right)^{\frac{R}{1-q}}-1 \right] 
 \ee
 and 
 \be
 S_{SM}(A) = \frac{k_B}{R} \left[\left(\sum_{i=1}^n (p_i^A)^{q}\right)^{\frac{R}{1-q}}-1 \right] .
 \ee
 Calculation of the RHS of (\ref{SMrule}) gives
 \bea
 && RHS =  \frac{k_B}{R} \left[\left(\sum_{i=1}^n (p_i^A)^{q}\right)^{\frac{R}{1-q}}-1 \right]  + \frac{k_B}{R} \left[\left(\sum_{j=1}^m (p_j^B)^{q}\right)^{\frac{R}{1-q}}-1 \right]  + \frac{\Upsilon}{k_B} 
 \frac{k_B}{R} \left[\left(\sum_{i=1}^n (p_i^A)^{q}\right)^{\frac{R}{1-q}}-1 \right]  \times \nonumber \\ 
&&  \frac{k_B}{R} \left[\left(\sum_{j=1}^m (p_j^B)^{q}\right)^{\frac{R}{1-q}}-1 \right] = \frac{k_B}{R} \left[ \left(\sum_{i=1}^n (p_i^A)^{q}\right)^{\frac{R}{1-q}} - 1 + \left(\sum_{j=1}^m (p_j^B)^{q}\right)^{\frac{R}{1-q}}  - 1 \right. \nonumber \\
&& \left. + \frac{\Upsilon}{R} \left(\sum_{i=1}^n (p_i^A)^{q}\right)^{\frac{R}{1-q}} \left(\sum_{j=1}^m (p_j^B)^{q}\right)^{\frac{R}{1-q}}  -  \frac{\Upsilon}{R} \left(\sum_{i=1}^n (p_i^A)^{q}\right)^{\frac{R}{1-q}}  -
 \frac{\Upsilon}{R} \left(\sum_{j=1}^m (p_j^B)^{q}\right)^{\frac{R}{1-q}} +  \frac{\Upsilon}{R}   \right]  \nonumber \\
&& =  \frac{k_B}{R} \left[\left(\sum_{i=1}^n \sum_{j=1}^m (p_{i}^{A})^q (p_j^B)^{q}\right)^{\frac{R}{1-q}}-1 \right]  = \frac{k_B}{R} \left[\left(\sum_{i=1}^n \sum_{j=1}^m (p_{ij}^{A+B})^{q}\right)^{\frac{R}{1-q}}-1 \right]  = LHS ,
 \eea
 where we have taken $\Upsilon = R$, and applied (\ref{multi_sum}).






\reftitle{References}







\bibliography{Entropy2024}{}
\end{document}